\documentclass[a4paper,12pt]{article}\pdfoutput=1
\usepackage{graphicx, rotating}
\usepackage{slashed}

\ifx\pdfoutput\undefined
\usepackage[dvips,bookmarks]{hyperref}	
\else
\usepackage{hyperref}	
\fi
\hypersetup{colorlinks,bookmarksopen,bookmarksnumbered,citecolor=verdes,
linkcolor=blus,pdfstartview=FitH,urlcolor=rossos}
\def\hhref#1{\href{http://arxiv.org/abs/#1}{#1}} 

\newcommand{\cm}{\,{\rm cm}}

\usepackage{multicol}
\usepackage{color}
\definecolor{rosso}{cmyk}{0,1,1,0.4}
\definecolor{rossos}{cmyk}{0,1,1,0.55}
\definecolor{rossoc}{cmyk}{0,1,1,0.2}
\definecolor{blu}{cmyk}{1,1,0,0.3}
\definecolor{blus}{cmyk}{1,1,0,0.6}
\definecolor{bluc}{cmyk}{1,1,0,0.1}
\definecolor{verde}{cmyk}{0.92,0,0.59,0.25}
\definecolor{verdec}{cmyk}{0.92,0,0.59,0.15}
\definecolor{verdes}{cmyk}{0.92,0,0.59,0.4}

\font\tenrsfs=rsfs10 at 12pt
\font\sevenrsfs=rsfs7
\font\fiversfs=rsfs5
\newfam\rsfsfam
\textfont\rsfsfam=\tenrsfs
\scriptfont\rsfsfam=\sevenrsfs
\scriptscriptfont\rsfsfam=\fiversfs
\def\mathscr#1{{\fam\rsfsfam\relax#1}}

\def\lsim{\ \rlap{\raise 3pt \hbox{$<$}}{\lower 3pt \hbox{$\sim$}}\ }
\def\gsim{\ \rlap{\raise 3pt \hbox{$>$}}{\lower 3pt \hbox{$\sim$}}\ }

\newcommand{\fig}[1]{~\ref{fig:#1}}
\newcommand{\eq}[1]{~{\rm (\ref{eq:#1})}}

\newcommand{\GeV}{\,{\rm GeV}}
\newcommand{\TeV}{\,{\rm TeV}}

\def\circa#1{\,\raise.3ex\hbox{$#1$\kern-.75em\lower1ex\hbox{$\sim$}}\,}

\newcommand{\DM}{{\rm DM}}

\newcommand{\beq}{\begin{equation}}
\newcommand{\eeq}{\end{equation}}

\def\circa#1{\,\raise.3ex\hbox{$#1$\kern-.75em\lower1ex\hbox{$\sim$}}\,}
\makeatletter

\def\art{\@ifnextchar[{\eart}{\oart}}
\def\eart[#1]#2#3#4#5#6{{\rm #2}, {#3 #4} {\rm (#6) #5} [{\hhref{#1}}]}
\def\hepart[#1]#2{{\rm #2, \hhref{#1}}}
\newcommand{\oart}[5]{{\rm #1}, {#2 #3} {\rm (#5) #4}}

\newcounter{alphaequation}[equation]
\def\thealphaequation{\theequation\hbox to
0.6em{\hfil\alph{alphaequation}\hfil}}
\def\eqnsystem#1{
\def\@eqnnum{{\rm (\thealphaequation)}}
\def\@@eqncr{\let\@tempa\relax \ifcase\@eqcnt \def\@tempa{& & &} \or
  \def\@tempa{& &}\or \def\@tempa{&}\fi\@tempa
  \if@eqnsw\@eqnnum\refstepcounter{alphaequation}\fi
\global\@eqnswtrue\global\@eqcnt=0\cr}
\refstepcounter{equation} \let\@currentlabel\theequation \def\@tempb{#1}
\ifx\@tempb\empty\else\label{#1}\fi
\refstepcounter{alphaequation}
\let\@currentlabel\thealphaequation
\global\@eqnswtrue\global\@eqcnt=0 \tabskip\@centering\let\\=\@eqncr
$$\halign to \displaywidth\bgroup \@eqnsel\hskip\@centering
$\displaystyle\tabskip\z@{##}$&\global\@eqcnt\@ne
\hskip2\arraycolsep\hfil${##}$\hfil& \global\@eqcnt\tw@\hskip2\arraycolsep
$\displaystyle\tabskip\z@{##}$\hfil
\tabskip\@centering&\llap{##}\tabskip\z@\cr}
\def\endeqnsystem{\@@eqncr\egroup$$\global\@ignoretrue} \makeatother

\newcommand{\SU}{{\rm SU}}

\begin{document}

\begin{center}
IFUP-TH/2008-37\hfill ~~

\bigskip\bigskip

{\huge\bf\color{magenta}
Decaying Dark Matter\\[5mm]
can explain
the $e^\pm$ excesses}

\medskip
\bigskip\color{black}\vspace{0.6cm}
{
{\large\bf Enrico Nardi$^{a,b}$, Francesco Sannino$^c$, Alessandro Strumia$^d$}
}
\\[7mm]
{\it $^a$  INFN-Laboratori Nazionali di Frascati, C.P. 13, I-00044 Frascati, Italia}
\\[2mm]
{\it $^b$  Instituto de F{\'i}sica, Universidad de Antioquia, A.A.1226 Medell{\'i}n, Colombia}
 \\[2mm]
{\it $^c$Centre for High Energy Physics, University of Southern Denmark,  Denmark}
\\[2mm]
{\it $^d$ Dipartimento di Fisica dell' Universit{\`a} di Pisa and INFN, Italia}

\bigskip\bigskip\bigskip

\thispagestyle{empty}

{\large
\centerline{\large\bf Abstract}

\begin{quote}
  \large PAMELA and ATIC recently reported excesses in $e^\pm$ cosmic rays.
  Since the interpretation in terms of DM annihilations was found to be not easily
  compatible with constraints from photon observations, we consider the 
   DM decay hypothesis and find that it can explain the
  $e^\pm$ excesses compatibly with all constraints, and can be tested by
  dedicated HESS observations of the Galactic Ridge.  ATIC data
  indicate a DM mass of about 2 TeV: this mass naturally implies the
  observed DM abundance relative to ordinary matter if DM
  is a quasi-stable composite particle with a baryon-like matter asymmetry.
  Technicolor naturally yields these type of candidates.
\end{quote}}

\end{center}


\newpage

\section{Introduction}
The recent observations of Cosmic Ray $e^\pm$ spectra
by PAMELA~\cite{PAMELA} and ATIC~\cite{ATIC-2}
indicate an excess, compatibly with other recent results~\cite{PAMELApbar,HESSe+e-}.
Unless it is due to pulsars or some other astrophysical source,
this excess could be the first  non-gravitational manifestation of Dark Matter.  
However, the interpretation in terms of DM annihilations into
SM particles is possible only for a restricted class of DM models~\cite{CKRS,MDM3,post-pamela,Nima}
leading to an unobserved excess of gamma or radio photons, unless the 
DM density profile is significantly less steep than the NFW profile
at galactocentric radii below 100 pc~\cite{BCST}.

In this paper we show that the interpretation in terms of DM
decays~\cite{DMdecay} circumvents these issues.  In section~2 we compute the
`halo functions' that encode the astrophysical information relevant for
understanding the energy spectra of $e^\pm$ and $\bar p$ produced by DM
decays.  In section~3 we propose a model-independent interpretation of the
PAMELA and ATIC observations in terms of DM decays, taking into account
constraints from $\bar p$ data.  In section~4 we show that the models that can
explain the PAMELA and ATIC excesses are compatible with photon observations
at gamma and radio frequencies, and single out the future observations capable
of testing the decaying DM interpretation.  
We will find that:
\begin{itemize}
\item When interpreted in terms of decaying DM, PAMELA $e^+$ and ATIC $e^+ +
  e^-$ data can be accounted for by a DM particle with mass around $2\,$TeV 
  and lifetime $\approx 10^{26}\,$s.  Decay modes yielding hadrons are
  constrained by PAMELA $\bar p$ data and should not exceed $10\%$.
\end{itemize}
In section~5 we explore the possible theoretical realizations of such scenario, finding that:
\begin{itemize}
\item If DM particles carry an anomalous global charge $B'$, the observed
DM abundance, $\Omega_{\rm DM}/\Omega_{B}\approx 5$,
is naturally obtained for $M_{\rm  DM}\approx 2\,$TeV.

\item  The above situation can be realized assuming that DM is a composite state,
suggesting a connection with technicolor:
if the hypothetical DM is the lightest 
state carrying  techni-baryon number
$B'$, a naive rescaling of QCD  suggests a
$M_{\rm DM}\approx 2\,$TeV mass.

\item If $B'$ is violated by dimension-6 higher dimensional operators suppressed by 
some `GUT scale' $\approx 10^{16}\,$GeV, DM is quasi-stable and
the desired DM life-time is naturally obtained.
\end{itemize}
Section~6 summarizes our results.

\medskip

\begin{figure}[t]
\begin{center}
 \includegraphics[width=0.45\textwidth]{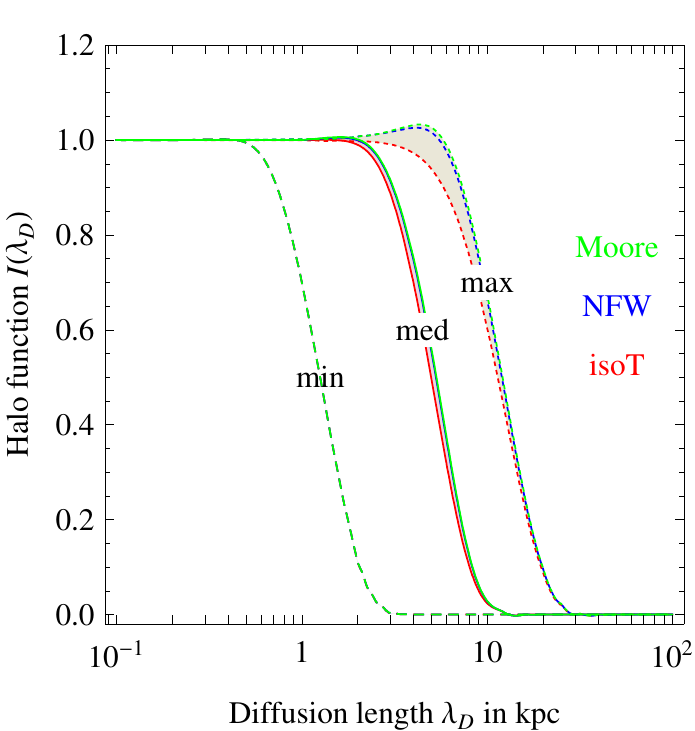}\qquad
\includegraphics[width=0.45\textwidth]{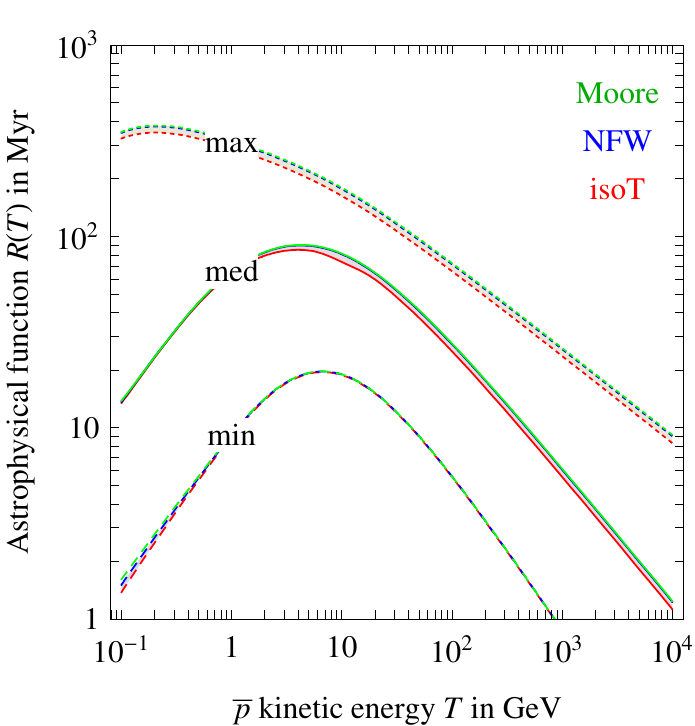}
  \caption{\em\label{fig:HaloI} Left: The uncertain `halo function'
    $I(\lambda_D)$ of eq.\eq{fluxpositrons} that encodes the astrophysics of
    DM decays into positrons and their propagation to the Earth. The
    diffusion length is related to energy losses as in eq.\eq{lambdaD}.
    Right: The $\bar p$ astrophysical function $R(T)$ of eq.\eq{RT}, computed
    under different assumptions.  In both cases, the dashed (solid) [dotted]
    bands assumes the {\rm min (med) [max]} propagation configuration of
    eq.\eq{proparampositrons} and eq.\eq{proparam} respectively.  Each band
    contains 3 lines, that correspond to the isothermal (red lower lines), NFW
    (blue middle lines) and Moore (green upper lines) DM density profiles.  }
\label{default}
\end{center}
\end{figure}

\section{Astrophysics}
The computation of the fluxes of SM particles $k=\{e^+,\bar p,\ldots\}$ from
DM decays is analogous to the case of DM annihilations: in brief one needs to
replace the ``annihilations'' source term
\beq 
\label{eq:Qann} Q_k(\vec x, E) = \frac{1}{2} \left(\frac{\rho(\vec
    x)}{M_{\rm DM}}\right)^2 \langle \sigma v\rangle \frac{dN_k}{dE}\qquad
\hbox{(DM annihilations)} 
\eeq
with
\beq \label{eq:Qdec}
Q_k(\vec x,E) =\frac{\rho(\vec x)}{M_{\rm DM}} \Gamma \frac{dN_{k}}{dE}
\qquad
\hbox{(DM decays)}.
\eeq
where $dN_k/dE$ is the spectrum of particles $k$ produced by one decay or
annihilation, $\rho(\vec x)$ the DM energy density at $\vec x$ and $M_{\rm DM}$
its mass.  Furthermore DM sub-halos can enhance the DM annihilation signals by
a potentially large `boost factor' $B_k\ge 1$; for DM decays one expects a
boost factor negligibly different from unity.

We briefly summarize aspects of $e^\pm$ and $p^\pm$ in our galaxy in order to
present the astrophysical functions, plotted in fig.\fig{HaloI}, that allow to
compute their fluxes at Earth in the generic case.

\subsection{Positron propagation}
The flux per unit energy of ultra-relativistic positrons is given by
$\Phi_{e^+}(t,\vec x,E) = f/4\pi$ where the positron number density per unit
energy, $f(t,\vec x,E)= dN_{e^+}/dE$, obeys the stationary diffusion-loss
equation:
\beq \label{eq:diffeq}-K(E)\cdot \nabla^2f - 
\frac{\partial}{\partial E}\left(b(E) f \right) = Q_e 
\eeq
with diffusion coefficient $K(E)=K_0 (E/\GeV)^\delta$ and energy loss
coefficient $b(E)=E^2/(\GeV\cdot \tau_E)$ with $\tau_E = 10^{16}\,{\rm s}$.
They respectively describe transport through the turbulent magnetic fields and
energy loss due to synchrotron radiation and inverse Compton scattering on
galactic photons.  Eq.~(\ref{eq:diffeq}) is solved in a diffusive region with
the shape of a solid flat cylinder that sandwiches the galactic plane, with
height $2L$ in the $z$ direction and radius $R=20\,{\rm kpc}$ in the $r$
direction. The location of the solar system
corresponds to $\vec x = (r_{\odot}, z_{\odot}) = ((8.5\pm0.5)\, {\rm kpc},
0)$.  The boundary conditions impose that the positron density $f$ vanishes on
the surface of the diffusive cylinder, outside of which turbulent magnetic
fields can be neglected so that positrons freely propagate and escape.  The
values of the propagation parameters $\delta$, $K_0$ and $L$ are deduced from
a variety of cosmic ray data and modelizations. We adopt the sets discussed
in~\cite{FornengoDec2007}:
\beq \begin{tabular}{lccc}
Model  & $\delta$ & $K_0$ in kpc$^2$/Myr & $L$ in kpc  \\
\hline 
min  & 0.55 &  0.00595 & 1 \\
med  & 0.70 &  0.0112 & 4  \\
max   & 0.46 &  0.0765 & 15 
\end{tabular}\label{eq:proparampositrons}
\eeq

Analogously to the DM annihilation
case~\cite{MDM3,FornengoDec2007,HisanoAntiparticles}, the solution for the
positron flux at Earth can be written in a useful semi-analytical form
\beq 
\Phi_{e^+}(E,\vec r_{\odot}) = \frac{  \Gamma}{4\pi b(E)}
\frac{\rho_\odot}{M_{\rm DM}}  \int_{E}^{M_{\rm DM}}  
dE'~\frac{dN_{e^+}}{dE'}\cdot  I \left(\lambda_D(E,E')\right)
\label{eq:fluxpositrons}
\eeq
where $\lambda_D(E,E')$ is the diffusion length from energy $E'$ to energy
$E$: 
\beq 
\lambda_D^2= 4K_0 \tau_E \left[\frac{(E'/\GeV)^{\delta-1} -
    (E/\GeV)^{\delta-1}}{\delta-1}\right]
\label{eq:lambdaD}
\eeq
the adimensional `halo function for DM decays' $I(\lambda_D)$ fully encodes
the galactic astrophysics and is independent on the particle physics model.
Its possible shapes are plotted in fig.\fig{HaloI} for the set of DM density
profiles (isothermal, Navarro, Frank and White (NFW)~\cite{Navarro:1995iw} and
Moore~\cite{Moore:1997sg}) and $e^+$ propagation models that we consider.  We
see that, unlike in the case of DM annihilations, $I$ is negligibly affected
by the uncertainty in the DM density profile, and DM decays in the center of
the galaxy (at 8.5 kpc from us) are never dominant.

\subsection{Antiproton propagation}
The propagation of anti-protons through the galaxy is described by a diffusion
equation analogous to the one for positrons.  Again, the number density of
anti-protons per unit energy $f(t,\vec x,T) = dN_{\bar p}/dT$ vanishes on the
surface of the cylinder at $z=\pm L$ and $r=R$. $T=E-m_p$ is the $\bar p$
kinetic energy, conveniently used instead of the total energy $E$. Since
$m_p\gg m_e$ we can neglect the energy loss term, and the diffusion equation
for $f$ is
\beq 
\label{eq:diffeqp}
- K(T)\cdot \nabla^2f + \frac{\partial}{\partial z}\left( {\rm sign}(z)\, f\,
  V_{\rm conv} \right) = Q_p-2h\, \delta(z)\, \Gamma_{\rm ann} f 
\eeq
where:
\begin{itemize}

\item[-] The diffusion term can again be written as $K(T) = K_0 \beta \,
  (p/\GeV)^\delta$, where $p$ and $\beta $ are the antiproton momentum and
  velocity. $\delta$ and $K_0$ are given in table~\eq{proparam}.

\item[-] The $V_{\rm conv}$ term corresponds to a convective wind, assumed to
  be constant and directed outward from the galactic plane, that tends to push
  away $\bar p$ with energy $T \circa{<}10\, m_p$. Its value is given in\eq{proparam}.

\item[-] The last term in eq.\eq{diffeqp} describes the annihilations of $\bar
  p$ on interstellar protons in the galactic plane (with a thickness of
  $h=0.1\,{\rm kpc} \ll L$) with rate $\Gamma_{\rm ann} = (n_{\rm H} + 4^{2/3}
  n_{\rm He}) \sigma^{\rm ann}_{p\bar{p}} v_{\bar{p}}$, where $n_{\rm
    H}\approx 1/{\rm cm}^3$ is the hydrogen density, $n_{\rm He}\approx 0.07\,
  n_{\rm H}$ is the Helium density (the factor $4^{2/3}$ accounting for the
  different geometrical cross section in an effective way) and $ \sigma^{\rm
    ann}_{p\bar{p}}$ is given by \cite{HisanoAntiparticles} 
\beq 
\sigma_{p\bar p}^{\rm ann} = \left\{
\begin{array}{ll}
661\, (1+0.0115\, T^{-0.774} - 0.984\, T^{0.0151})\ {\rm mbarn}, 
& {\rm for}\ T < 15.5\, {\rm GeV} \\ 
36\, T^{-0.5}\ {\rm mbarn}, 
& {\rm for}\ T \geq 15.5\, {\rm GeV}
\end{array}
 \right. .
\label{eq:sigmaann}
\eeq
\item[-] We neglect ``tertiary $\bar p$'', i.e.\ non-annihilating $\bar p$
  interactions on the matter in the galactic disk.
\end{itemize}

The set of propagation parameters that we adopt in the case for anti-protons
has been deduced in \cite{DonatoPRD69}:
\beq \begin{tabular}{ccccc}
Model  & $\delta$ & $K_0$ in kpc$^2$/Myr & $L$ in kpc & $V_{\rm conv}$ in km/s \\
\hline 
min  & 0.85 &  0.0016 & 1  &13.5\\
med  & 0.70 &  0.0112 & 4  & 12 \\
max  & 0.46 &  0.0765 & 15 &5
\end{tabular}\label{eq:proparam}
\eeq

The solution for the antiproton flux at the position of the Earth $ \Phi_{\bar
  p}(T,\vec r_\odot) = v_{\bar p}/(4\pi) f $ can be written as
\beq
\Phi_{\bar p}(T,\vec r_\odot) = \Gamma \frac{v_{\bar p}}{4\pi}  
\frac{\rho_\odot}{M_{\rm DM}} R(T) \frac{dN^k_{\bar p}}{dT}, 
\label{eq:RT}
\eeq
where $\rho_\odot \equiv \rho(\vec r_\odot)$.  The `halo function for DM
decays' $R(T)$ encodes all the astrophysics and is plotted in fig.\fig{HaloI}
for the halo and propagation models that we consider.  It depends negligibly
on the DM density profile, unlike for the analogous function relevant for DM
annihilations, plotted in fig.~4 of~\cite{MDM3}.

\begin{figure}[t]
\begin{center}
  $$ \includegraphics[width=0.65\textwidth]{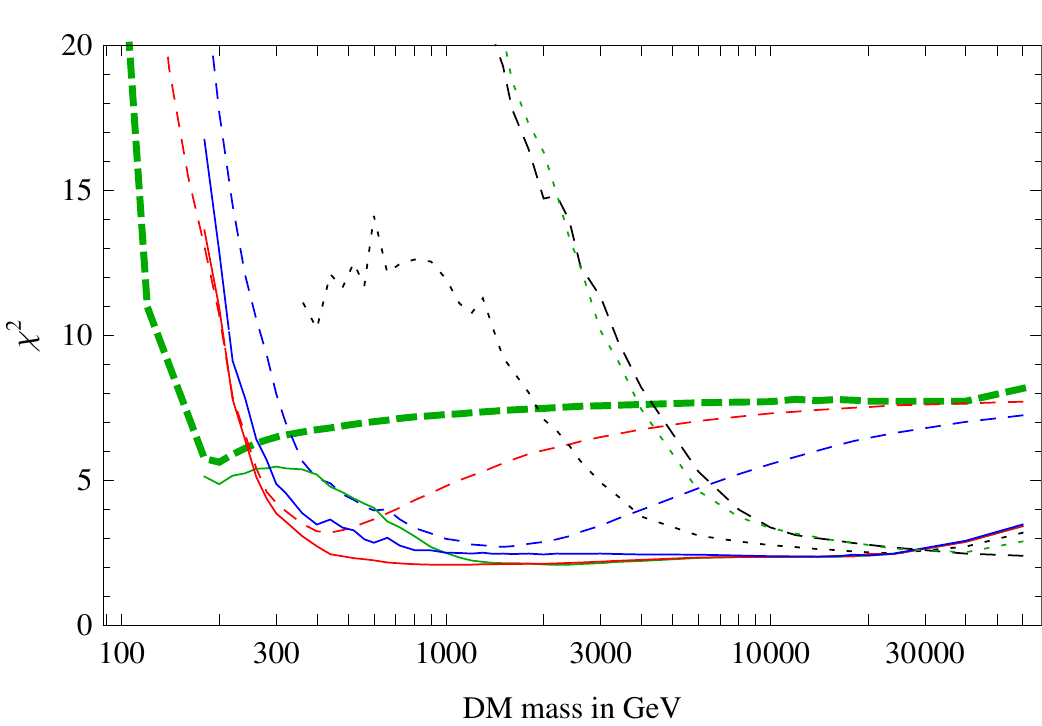}\qquad
 \raisebox{10mm}{\includegraphics[height=5cm]{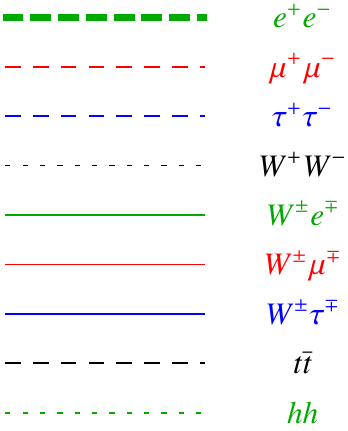}}$$
\caption{\em A fit of the DM decays indicated in the legend
 to the 
 PAMELA positron data.
\label{fig:fite}}
\end{center}
\end{figure}

  \begin{figure}[t]
  \begin{center}
 $$\includegraphics[width=0.65\textwidth]{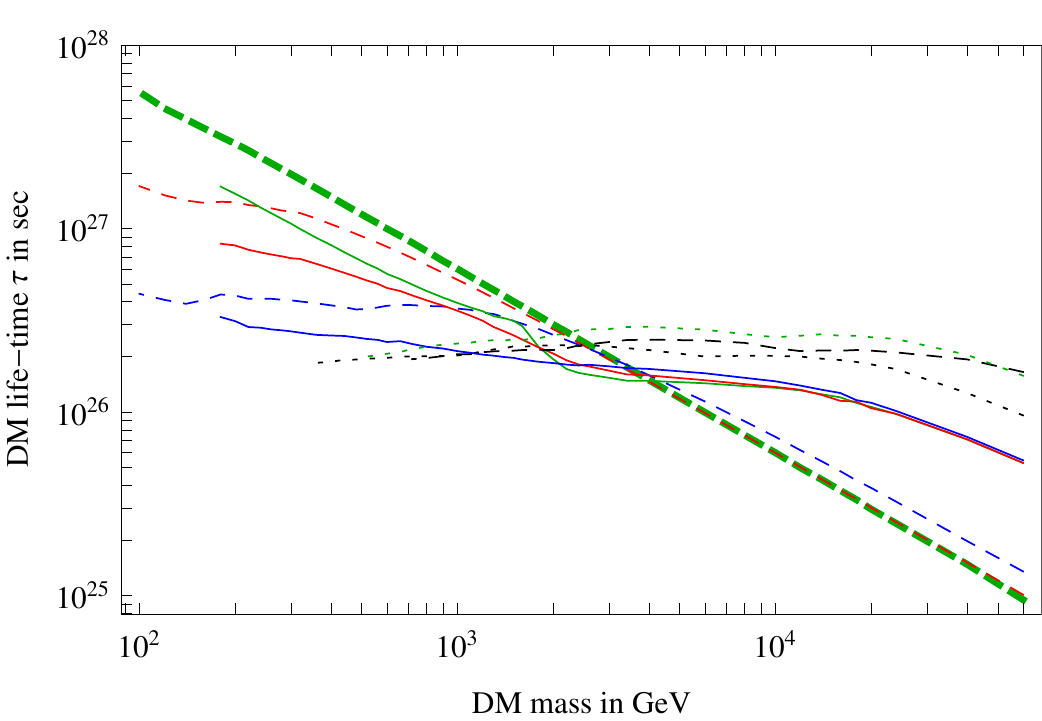}\qquad
 \raisebox{14mm}{\includegraphics[height=5cm]{legenda}}$$
  \caption{\em Best-fit values of the DM life-time suggested by the PAMELA excess,
  for the DM decay modes indicated on the legend.
  \label{fig:fitBooste}}\end{center}
  \end{figure}

  \begin{figure}[t]\begin{center}
$$\includegraphics[width=0.65\textwidth]{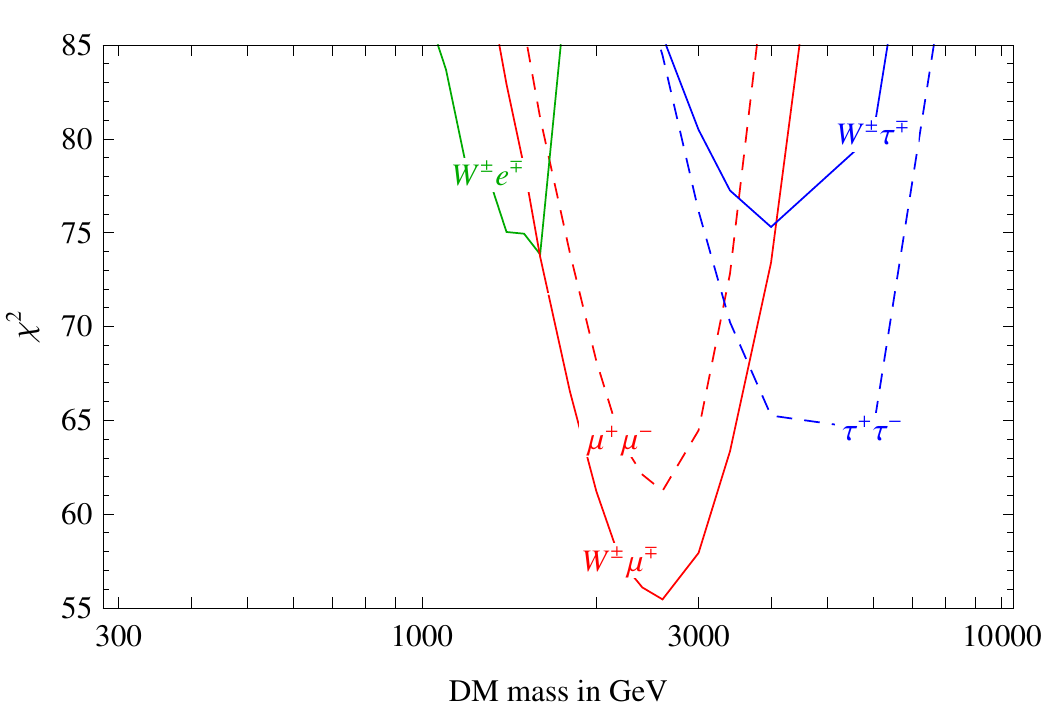}$$
  \caption{\em Combined fit of  positron data from PAMELA and of $e^++e^-$ data
  from ATIC, BBP-BETS, EC, HESS.
  \label{fig:fitee}}
  \end{center}
  \end{figure}

\begin{figure}[p]\begin{center}
$$\includegraphics[width=\textwidth]{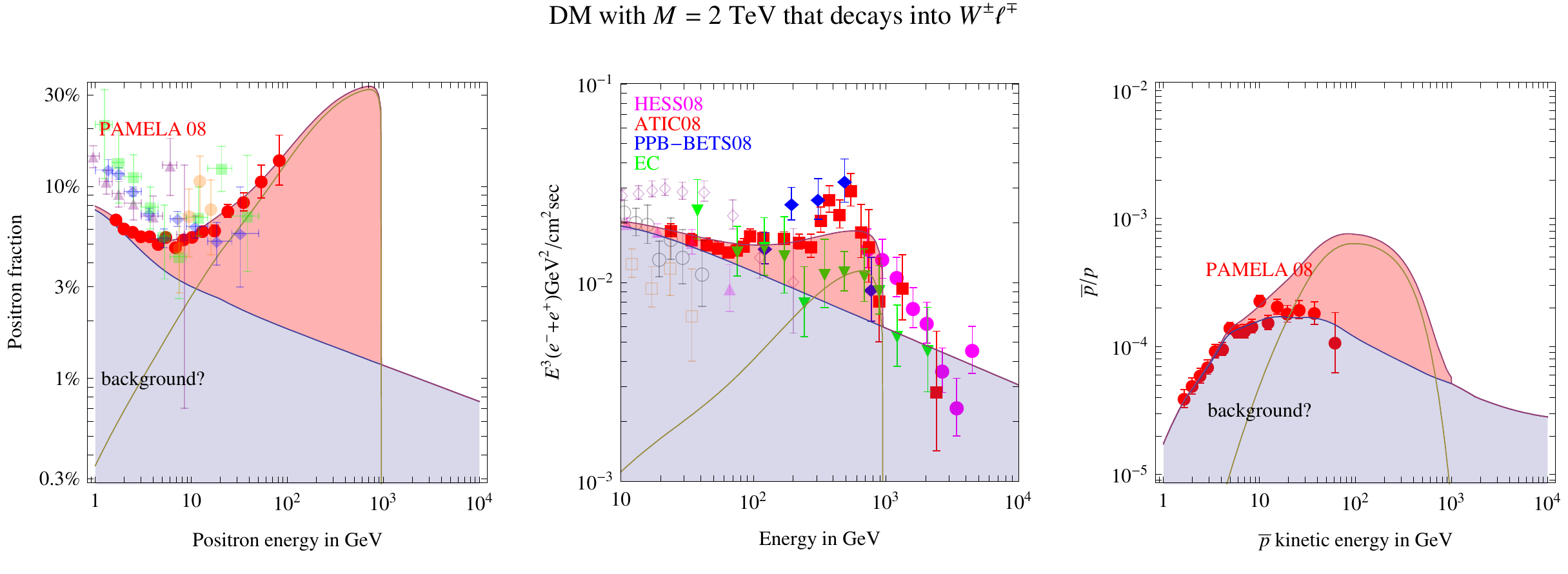}$$
$$\includegraphics[width=\textwidth]{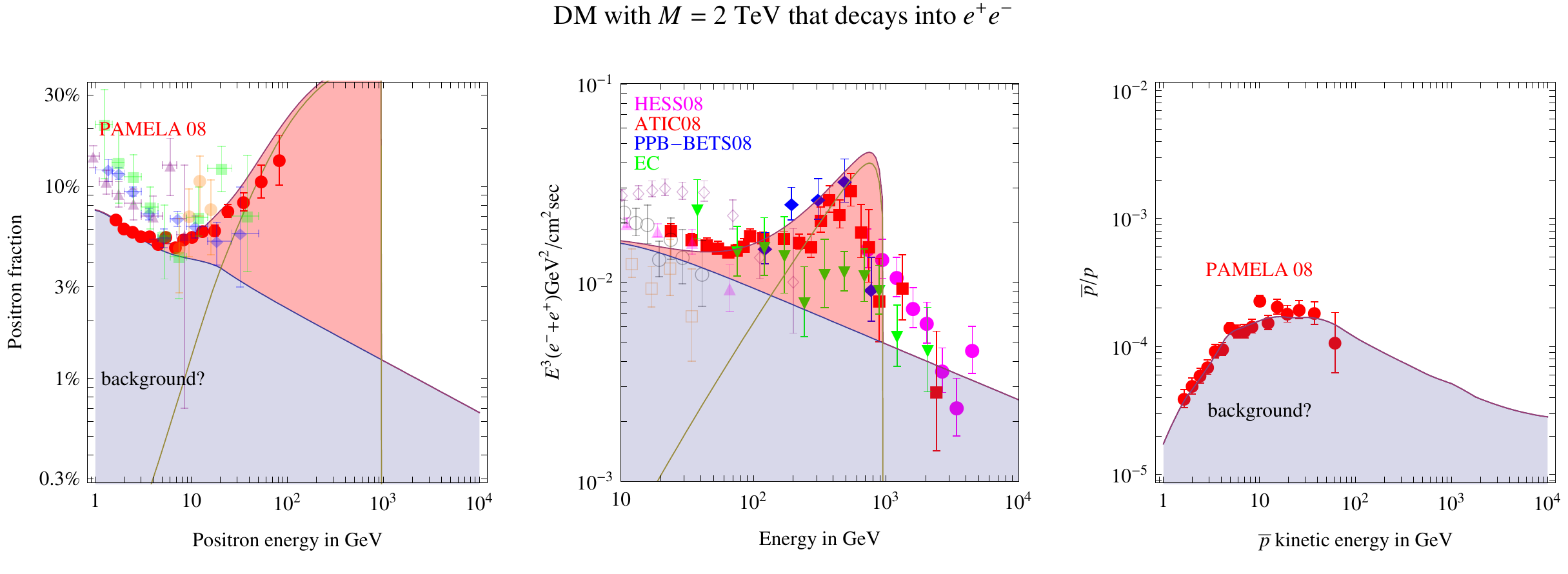}$$
$$\includegraphics[width=\textwidth]{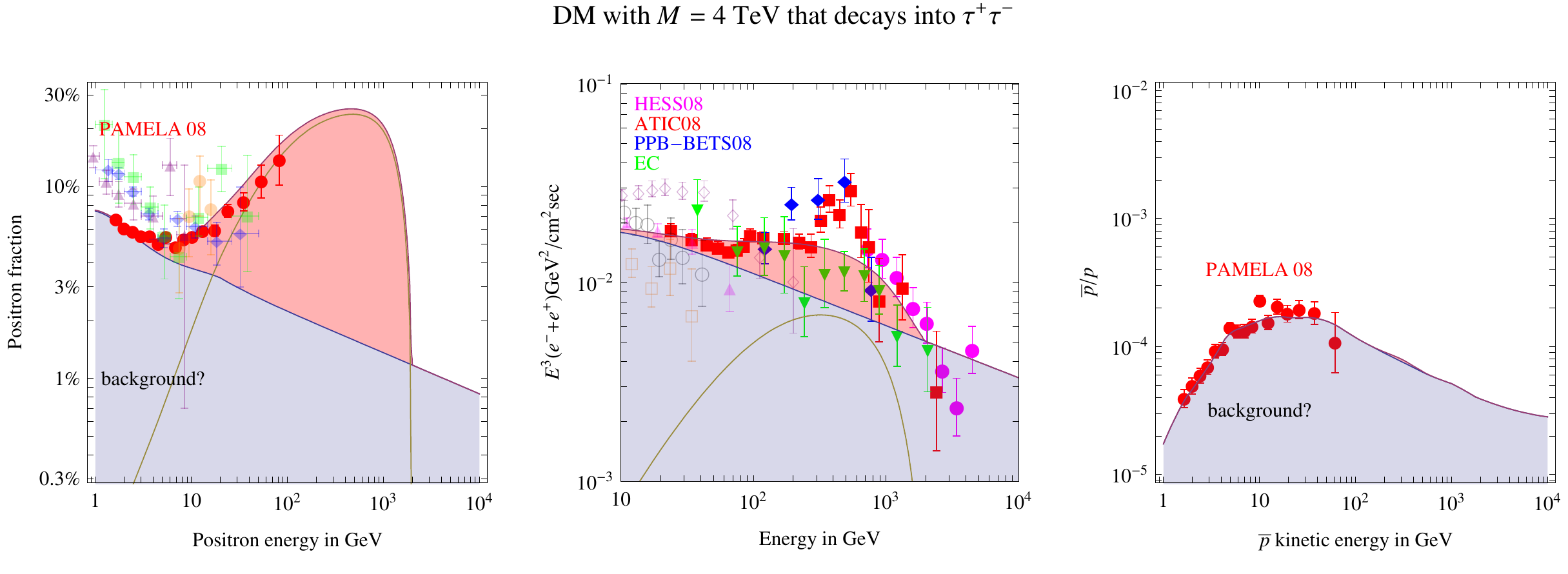}$$
\caption{\em Examples of fits of $e^+$ (left), $e^++e^-$ (center), $\bar{p}$
  (right) CR data, for DM decay modes into $W^\pm \ell^\mp$
  (that leads to an unseen $\bar p$ excess), $e^+e^-$
 (that leads to a too sharp peak),  $\tau^+\tau^-$
 (that leads to peak less sharp than suggested by ATIC data).
A combination of the last two possibilities, and the intermediate $\DM\to \mu^+\mu^-$ decay
gives the optimal fit.
  \label{fig:samples}}
\end{center}
\end{figure}

  \begin{figure}[t]\begin{center}
$$\includegraphics[width=0.65\textwidth]{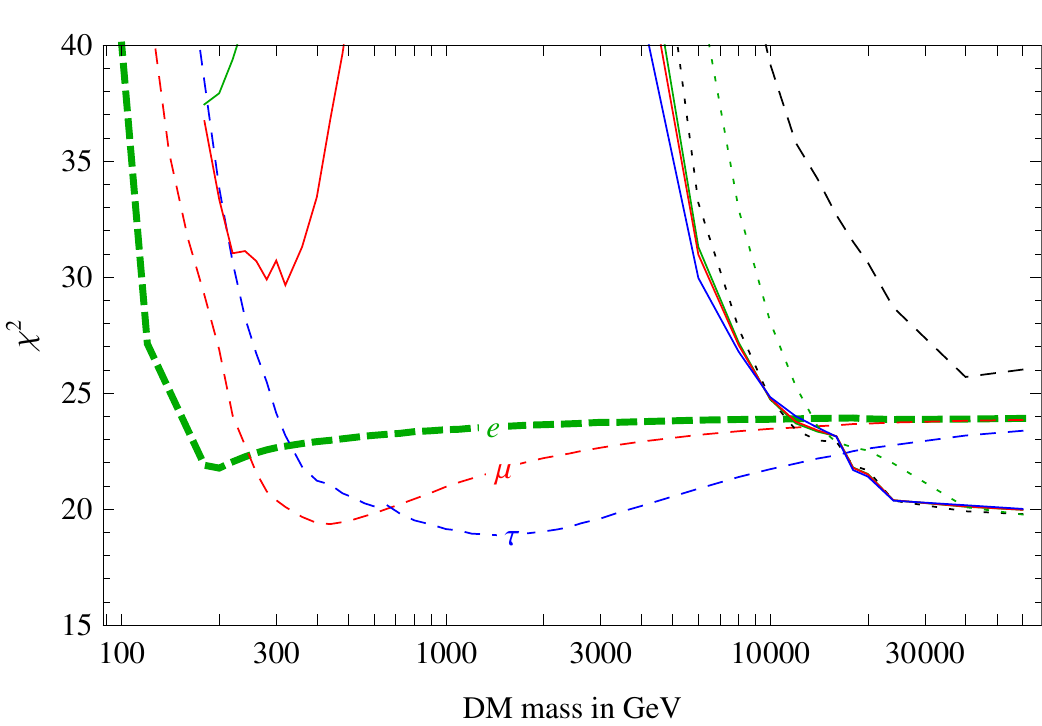}\qquad
\raisebox{14mm}{\includegraphics[height=5cm]{legenda}}$$
  \caption{\em Combined fit of different DM decay channels to the 
   PAMELA positron and PAMELA anti-proton data,
  assuming equal  propagation models for $e^\pm$ and $\bar p$.
  \label{fig:fitep}}
  \end{center}
  \end{figure}

\section{The PAMELA and ATIC data as DM decays}
With the goal of performing a model-independent analysis, we
delineate the generic features implied by known physics.  We recall that
non-relativistic DM annihilations are equivalent to the decay of the $s$-wave
DM~DM two-body state with mass $2M_{\rm DM}$, that can only have spin 0, 1 or 2,
if DM is a weakly-interacting particle~\cite{CKRS}.

The decay of a DM is less constrained: since DM could be coupled only
gravitationally to SM particles, its spin can be anything: 0, 1/2, 1, 3/2, 2,
etc.  Furthermore, since DM decay must be slow, decays into two SM particles
do not need to dominate over multi-body decays; decay rates suppressed by
helicity factors such as $(m_e/M_{\rm DM})^2$ are still phenomenologically interesting;
DM decays might involve new particles (e.g.\ gravitino decays into gluinos and
quarks).

At the light of the results of~\cite{CKRS} that single out decays into hard
leptons as phenomenologically promising we will study the following DM decay
modes.  For boson DM:
\beq 
\DM \to e^+ e^-,\mu^+\mu^-,\tau^+\tau^-, W^+ W^-, t \bar{t},hh
\eeq
where the $t\bar{t}$ and $hh$ modes are included as examples of hadronic modes.
For fermion DM we consider:
\beq 
\DM \to W^\pm e^\mp, W^\pm \mu^\mp,W^\pm \tau^\mp .
\eeq
3 body decays, such as $\DM \to \ell^+\ell^-\nu$ are also possible, but cannot
be computed in a model-independent way.

We fit the PAMELA and ATIC data taking into account as described
in~\cite{CKRS} the uncertainties on the astrophysical backgrounds, on the
$e^\pm $and $\bar p$ propagation, on the DM density profile, on the
experimental data.

\subsection{PAMELA $e^+$ data}

We start including only the PAMELA data about the positron fraction~\cite{PAMELA}, that
exhibit an excess above 10 GeV.  Fig.\fig{fite} shows the quality  of the best-fit as
function of the DM mass for the different decay modes we consider, and
fig.\fig{fitBooste} shows the corresponding best-fit values of the DM
life-time, that is $\tau \sim 10^{26}\,{\rm sec}$.
The left column of fig.\fig{samples} show three possible fits.

\subsection{PAMELA $e^+$ and $e^++e^-$ data}
Next, we explore the combined fit of PAMELA $e^+$ and the $e^++e^-$ data,
as measured by balloon experiments~\cite{ATIC-2} and at higher energies by HESS~\cite{HESSe+e-}.
We plot and fit the HESS data combining in quadrature a systematic $\pm 20\%$ systematic error
with a $\pm15\%$ energy-scale uncertainty with the statistical uncertainty on each data point~\cite{HESSe+e-}.
Fig.\fig{fitee} shows our results: the peak possibly present in the $e^++e^-$
energy spectrum, as measured by ATIC-2 and PPB-BETS, strongly restricts the DM
mass and the DM decay modes.  Like in the DM annihilation case, the
$\mu^+\mu^-$ and $\tau^+\tau^-$ modes provide good fits, now for a value of
the DM mass twice as large: $M\approx 3\TeV$.  Unlike in the DM annihilation
case, the $e^+ e^-$ mode does not provide a good fit, due to the following
detailed argument, that can be weakened invoking additional astrophysical
uncertainties.  The problem is that the $e^\pm$ line remains too sharp.
Indeed DM decays (unlike DM annihilations) are not strongly enhanced close to
the galactic center, so that such $e^+$ produced far from us and that
therefore reach us at the price of significant energy losses are now a minor
component.  Furthermore, the new $W^\pm \ell^\mp$ decay modes possible for
fermion DM also provide good fits for $e^+$ and $e^++e^-$ data for any $\ell$.
A similar result holds for the various 3-body decays $\DM\to\ell^+ \ell^-\nu$,
that we cannot compute in a model-independent way.

\subsection{PAMELA $e^+$ and $\bar p$ data}
Fig.\fig{fitep} shows the result of the fit of PAMELA $e^+$ and $\bar p$
PAMELA observations.  The $\bar p$ data, that do not show an excess with
respect to the expected astrophysical background, provide significant
constraints, disfavoring decay modes that lead to hadrons.  The situation with
the $\DM \to W^+ W^-, t\bar t, hh$ modes is similar to the DM annihilation
case~\cite{CKRS}: such modes are disfavored unless $M \circa{>} 10\TeV$.  We
see that the same conclusion applies to the new semi-leptonic decay modes
$W^\pm \ell^\mp$: choosing $M\sim 2\TeV$ in order to fit the $e^++e^-$ peak in
ATIC-2 data is disfavored by PAMELA $\bar p$ data.  This is illustrated in the
upper row of fig.\fig{samples}, where we summed over all leptons
$\ell=\{e,\mu,\tau\}$ as suggested by gauge invariance.  We recall that we
assumed equal propagation models for $e^+$ and $\bar p$, as controlled by the
thickness $L$ of the cylinder where turbulent magnetic fields lead to
diffusion of charged particles.  Our conclusions can be weakened assuming
different propagation models for $\bar p$ and $e^+$.  Unlike in the DM
annihilation case it is not possible to weaken these conclusions by assuming a
boost factor $B_{p}\ll B_{e}$.

\medskip

Summarizing, a good fit to all $e^+$, $e^++e^-$ and $\bar p$ CR data presently
available in terms of decaying DM requires decays into $\mu^+\mu^-$ or
$\tau^+\tau^-$, which are possible if DM is a boson.  A fermion DM can decay
into $\ell^+\ell^- \nu$, which can also provide a good global fit, although we
cannot compute the resulting $e^\pm$ energy spectra in a model-independent
way.  If decaying DM is only a fraction $f$ of all DM, our results still hold
up to the rescaling $\tau\to \tau/f$.

\section{Photons: bounds and signals}
DM decays unavoidably lead to photons: at $\gamma$ frequencies due to brehmstralung of charged particles,
and at radio frequencies due to synchrotron radiation emitted by $e^\pm$ in the galactic magnetic fields.
Here we compare the flux of photons produced by DM decays with the existing
bounds and we comment on future perspectives.  There are three main robust
constraints and we find that, unlike in the case of DM
annihilations~\cite{BCST}, they are satisfied even in the case of a NFW DM
density profile.

\subsection{HESS observation of the Galactic Center}
The differential flux of photons received from a given angular direction
$d\Omega$ is
\beq \frac{d \Phi_\gamma}{d\Omega\,dE} =\Gamma \frac{r_\odot}{4\pi}
\frac{\rho_\odot}{M} \frac{dN_\gamma}{dE} \frac{dJ}{d\Omega} ,\qquad
\frac{dJ}{d\Omega} = \int_{\rm line-of-sight} \frac{ds}{r_\odot}
\frac{\rho}{\rho_\odot} \eeq
where $dN_\gamma/dE$ is the photon spectrum produced in one DM decay: we here
include only the model-independent contribution due to brehmstralung of
charged particles.  We conservatively impose that such contribution does
not exceed at $3\sigma$ the observed flux in any data-point.
The adimensional quantity $J$ describes the uncertain
astrophysics.  Observing an angular region $\Delta \Omega=10^{-5}$ centered on
the GC, we find \beq \hbox{GC $J$ for DM decays} = \{ 5.75 , 28.9 ,
45.3\}\quad \hbox{for the \{isoT, NFW, Moore\} profiles}\eeq
for the quantity $J$ defined by $J\cdot\Delta \Omega = \int d\Omega\cdot
dJ/d\Omega$.  In the case of DM annihilations, one instead has $J \cdot
\Delta\Omega=\int d\Omega \int (\rho/\rho_\odot)^2 ds/r_\odot$ and, the
analogous factor $J$ acquires much larger values, such as $J=14700$ for the
NFW profile,
implying that the PAMELA anomaly is not compatible with bounds from $\gamma$
observations, unless DM is distributed with an isothermal-like profile and/or the
$e^\pm$ signal is enhanced by a large boost factor
$B_e\circa{>}10^2\gg B_\gamma$~\cite{BCST}.

On the contrary this problematic issue is not present for the DM decay
interpretation of the PAMELA excess, as shown by the blue continuous lines in
fig.s\fig{boundsNFW}.  

\subsection{HESS observation of the Galactic Ridge}
It is unclear which observation region gives the maximal sensitivity to
$\gamma$ rays from DM decays.  Observing the Galactic Center (GC) maximizes
the signal rate, especially for cusped profiles such as NFW, where the DM
density $\rho$ grows as $1/r$ as $r \to 0$, and especially for DM
annihilations, where the signal rate is proportional to $\rho^2$.  However,
the GC also has the highest background rate, as HESS observations suggest that
it is polluted by astrophysical sources of $\gamma$ rays.

For DM decays, the signal is proportional to $\rho$ rather than to $\rho^2$,
so the optimal strategy might be instead observing a larger region.  HESS
observed the Galactic Ridge (GR), defined as the region corresponding to
longitude $|\ell| < 0.8^\circ$ and latitude $|b|<0.3^\circ$, with a cone of
angle $0.1^\circ$ centered on the GC subtracted.

This HESS observation leads to the dominant constraint on DM decays.  Indeed
HESS finds comparable total $\gamma$ fluxes from the GC and the GR regions,
$d\Phi_{\rm GR}/dE_\gamma \sim 3d\Phi_{\rm CG}/dE_\gamma$, and we find
comparable values of $J$:
\beq \hbox{GR $J$ for DM decays} = \{ 5.55 , 20.5 ,
26.8\}\quad \hbox{for the \{isoT, NFW, Moore\} profiles}.
\eeq 
Therefore, since $\Delta \Omega_{\rm GR} \sim 30 \Delta\Omega_{\rm GC}$,
observations of the GR set a bound on the DM life-time $\tau$ which is about
10 times stronger than the GC bound.  The blue dot-dashed lines in
fig.\fig{boundsNFW} show our precise result.

The GR HESS bound is close to the value of the DM life-time $\tau$ suggested
by PAMELA/ATIC.  Such bound can presumably be improved by observing some
region away from the Galactic Plane, to be chosen trying to avoid the
astrophysical background.  Unfortunately, astrophysicists are interested in
astrophysical `backgrounds', so that the GR region observed by HESS, that
contains the galactic plane, is clearly not the optimal choice.

\subsection{HESS observation of Sagittarius Dwarf}

\begin{figure}
 \begin{center}
$$\includegraphics[width=0.99\textwidth]{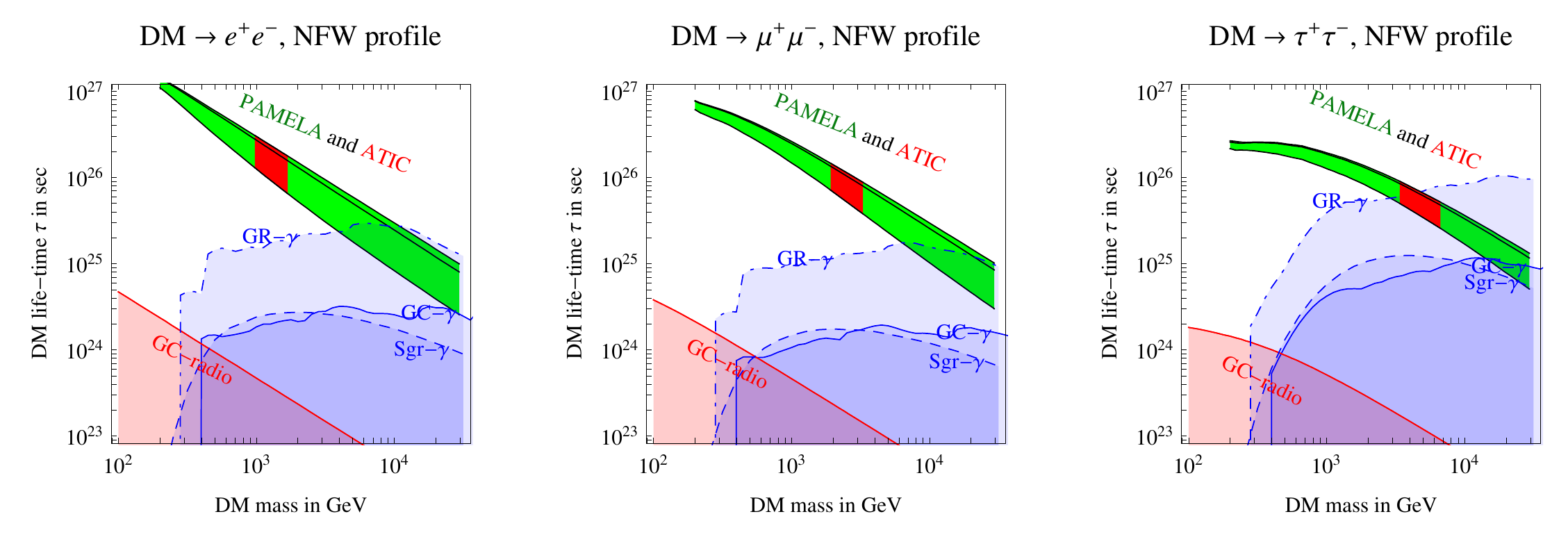}$$
$$\includegraphics[width=0.99\textwidth]{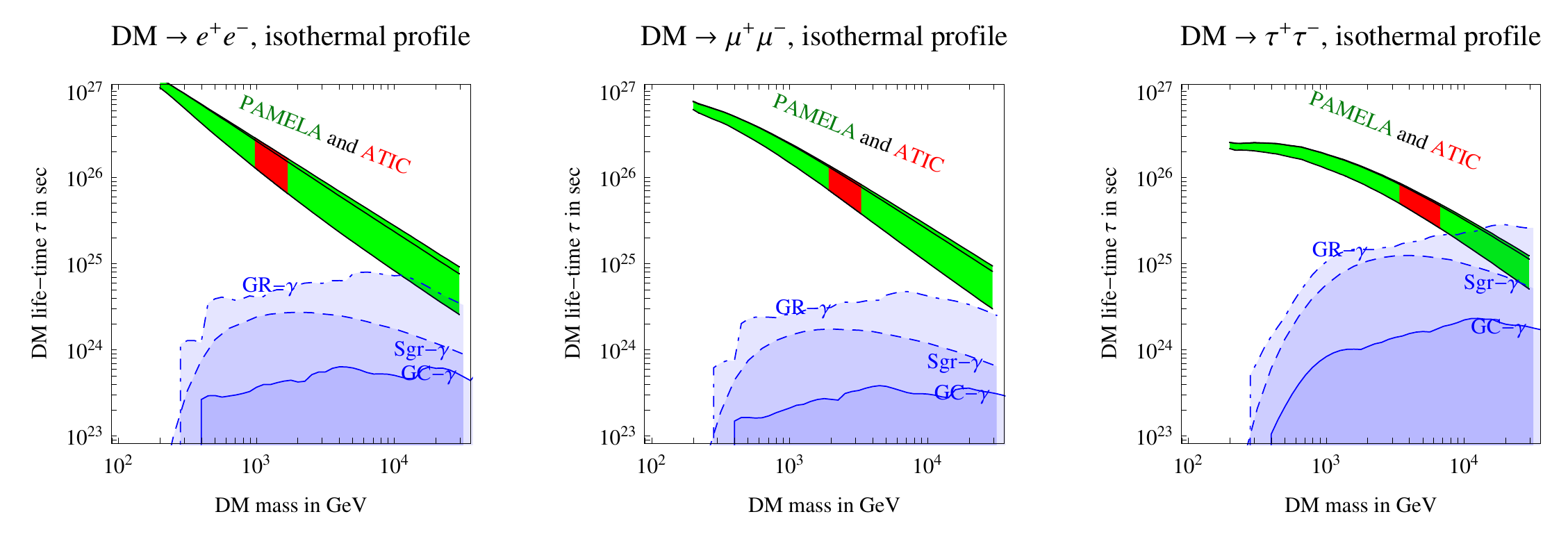}$$
 \caption{\em Assuming Dark Matter decays into $e^+ e^-$ (left column), $\mu^+\mu^-$
 (middle column), $\tau^+\tau^-$ (right column) we
 compare the region favored by 
 PAMELA (green band) and ATIC (red region), as computed for min/max/med $e^\pm$ galactic propagation,
   with the bounds from HESS gamma observations of the Galactic Center (blue continuous
  curves), of the Galactic Ridge (dot-dashed curves)  of Sagittarius Dwarf (blue dashed curves), 
 and with radiowave observations of the Galactic Center.
 We assumed NFW (upper row) and isothermal (lower row) DM density profiles.
  \label{fig:boundsNFW}}
  \end{center}
  \end{figure}

Dwarf spheroidals are among the most DM-dominated structures, so that they
allow to search for $\gamma$ ray signals of DM annihilations with minimal
astrophysical backgrounds.  HESS observed Sagittarius Dwarf at distance
$d=24\,{\rm kpc}$ from us for a time $T_{\rm obs}=11\,{\rm h}$ in a region
with angular size $\Delta \Omega=2~10^{-5}$, finding no $\gamma$ excess and
setting the bound $N_\gamma \circa{<}85$ at about $3\sigma$.  The $\gamma$
flux from DM decays is
\beq \frac{d\Phi_\gamma}{dE}=
\frac{\Gamma}{M}\frac{dN_\gamma}{dE}\frac{1}{4\pi d^2} \int_{\rm cone} dV
\rho, \qquad N_\gamma = T_{\rm obs} \int dE~A_{\rm eff}(E)
\frac{d\Phi_\gamma}{dE} \eeq
where $A_{\rm eff}(E)\sim 10^5\,{\rm m}^2$ is the effective area of
HESS~\cite{AeffHESS}.  The integral is the total DM mass $\mathscr{M}$
contained in the volume observed by HESS, and is equal to $\mathscr{M}\approx 5~10^6
M_\odot$ assuming a NFW or cored density profile~\cite{HESSSgrDwarf} for Sagittarius Dwarf.
Therefore, the bound on the DM life-time is:
\beq \tau> \frac{T_{\rm obs}}{N_\gamma^{\rm max}}
\frac{\mathscr{M}}{M}\frac{1}{4\pi d^2}\int dE~A_{\rm eff}(E)
\frac{dN_\gamma}{dE}.  \eeq
The resulting bound are plotted in fig.s\fig{boundsNFW} as blue dotted lines.

\subsection{Synchrotron radiation from the Galactic Center}
The $e^\pm$ produced by DM decays within the galactic magnetic field radiate
synchrotron radiation.  As the galactic magnetic fields are poorly known, in
order to discuss which bounds are robust we recall the basic physics.  One
expects $B\sim \mu{\rm G}$ that grows to higher values close to the galactic
center.  Such magnetic fields are intense enough that $e^\pm$ radiate an order
unity fraction of their energy into synchrotron radiation, as other energy
losses and diffusion are not the main phenomena.  Therefore the total energy
into synchrotron radiation can be robustly computed, and the value of the
magnetic field only determines how it is distributed in the frequency
spectrum.  We recall that the synchrotron power $W_\gamma$ radiated
orthogonally to the magnetic field $B$ by an $e^\pm$ with momentum $p$ is:
\beq\label{eq:syn}
\frac{dW_{\rm syn}}{d\nu}\approx \frac{2e^3B}{9m_e}
\delta(\frac{\nu}{\nu_{\rm syn}}-\frac{1}{3}) \qquad\hbox{where}\qquad
\nu_{\rm syn} = \frac{3eB p^2}{4\pi m_e^3}= 4.2\,{\rm MHz}
\frac{B}{\rm G} \left(\frac{p}{m_e}\right)^2.  
\eeq
Thereby a larger magnetic field and heavier DM mean that synchrotron radiation
extends up to higher energy.  In order to put a robust bound, we consider the
observation~\cite{Davies} of the GC in a region with angular size $4''$ at the
lowest radio-wave
frequency $\nu=0.408\,{\rm GHz}$, that implies the bound $S= (\nu \,
dW_{\rm syn}/d\nu)/(4\pi r_\odot^2) < 2~10^{-16}\,{\rm erg}/{\rm cm}^2{\rm
  sec}$.  This is the relevant bound, as the observed GC microwave spectrum is
harder than what DM decays can produce.  The region is large enough that we
can neglect $e^\pm$ diffusion and advection ($e^\pm$ falling in the central
black hole).  In this approximation the electron number density is given by
$n_e (r,p) \simeq {\dot E}^{-1}_{\rm syn} \int_E^\infty dE'~Q_e(r,E')$ with
$\dot E_{\rm syn} =e^4 B^2 p^2/9\pi m_e^4$ is the synchrotron energy loss of a
$e^\pm$ in a turbulent magnetic field.  So
\beq 
\nu \frac{dW_{\rm syn}}{d\nu} =\frac{\Gamma}{M}\int_{\rm cone}
dV~\rho~p~ N_e(p) 
\eeq
where the integral extends over the observed volume and $p$ is obtained from
eq.s\eq{syn} 
as $p=\sqrt{4\pi m_e^3\nu/B}=0.43\GeV (\nu/{\rm
  GHz})^{1/2}(B/{\rm mG})^{-1/2}$.  Here, $N_e(p)$ is the number of electrons
with energy above $p$ produced in one DM decay, and can be often approximated
as the total number of electrons produced in one DM decay.  As anticipated at
the beginning of this section, a lower $B$ leads to a higher synchrotron flux
at the low frequency we consider.

Observations are made at scales large enough that the 
dipolar structure of the GC can be neglected:
we model the GC as in~\cite{BCST} in `spherical-cow' approximation,
obtaining the bounds plotted  as red lines in fig.s\fig{boundsNFW}.
The radio-wave constraints turn out to be subdominant with respect to the $\gamma$-ray 
constraints previously computed.

\section{Theory}

In the DM annihilation case, model building is needed to invent models of DM
that annihilate dominantly into leptons with cross section larger than the one
suggested by cosmology.  Key ingredients of such models are the Sommerfeld
enhancement and possibly new vectors that decay into leptons~\cite{CKRS,MDM3,post-pamela}.

DM decay allows to circumvent such model-building issues.  It is very easy to
invent models of DM that decays dominantly to leptons, and small couplings to
leptons might be somehow connected to the small observed neutrino masses and
baryon/lepton cosmological asymmetries.  One example is analogous to type-II
see-saw~\cite{seesaw}: DM might be a scalar  $\SU(2)_L$ triplet $T$ with Lagrangian couplings $M^2 TT^*
+ \lambda_L TLL + M\lambda_T T HH$, so that it dominantly decays into leptons
if $\lambda_L\gg \lambda_T$.  
Alternatively, along the
lines of type-I or type-III see-saw~\cite{seesaw}, one can introduce fermion singlets or
triplets $N$ and an inert Higgs $H'$ with couplings (for simplicity we
consider the supersymmetric case and write the superpotential) $M N^2/2
+\lambda N L H' + \epsilon N^3$.  The neutral component of $H'$ could be DM
and decay into $L\bar{L}L$ with a slow rate suppressed by $(\epsilon/M)^2$.
Within supersymmetry, one can build models of gravitino DM that decays into leptons,
by assuming an appropriate sparticle spectrum and that $LLE$ operators dominate $R$-parity violation.
 DM might be some Froggatt-Nielsen-like field, and its slow decays into leptons might be related
to the quasi-conserved U(1)$_\ell$ flavor lepton numbers present in the lepton sector.

We now focus on a more interesting idea.

\subsection{Non-annihilating DM and the $\Omega_{\rm DM}/\Omega_B$ ratio}

A rather attractive scenario for decaying DM arises when one tries to relate
the DM and the baryon energy densities in order to  explain the ratio
$\Omega_{\rm DM}/\Omega_B\sim 5$~\cite{Komatsu:2008hk}.

We know that the amount of baryons in the Universe $\Omega_B\sim 0.04$ is
determined solely by the cosmic baryon asymmetry $n_B/n_\gamma \sim 6\times
10^{-10}$.  This is because the baryon-antibaryon annihilation cross section
is so large, that virtually all antibaryons annihilate away, and only the
contribution proportional to the asymmetry remains.  
This asymmetry can be
dynamically generated after inflation. In
contrast, we do not know if the DM density is determined by thermal freeze-out,
by an asymmetry, or by something else.
Thermal freeze-out needs a $\sigma v \approx 3~10^{-26}\cm^3/\sec$ of electroweak size,
suggesting a DM mass in the TeV range. If $\Omega_{\rm DM}$ is determined by thermal freeze-out, its proximity to $\Omega_B$
is just a fortuitous coincidence and is left unexplained.

If instead $\Omega_{\rm DM}\sim \Omega_B$ is not accidental, then the theoretical
challenge is to define a consistent scenario in which the two energy densities
are related.  Since $\Omega_B$ is a result of an asymmetry, then relating the
amount of DM to the amount of baryon matter can very well imply that
$\Omega_{\rm DM} $ is related to the same asymmetry that determines $\Omega_B$.
Such a condition is straightforwardly realized if the asymmetry for the DM
particles is fed in by the non-perturbative electroweak sphaleron transitions,
that at temperatures much larger than the temperature $T_*$ of the electroweak
phase transition (EWPT) equilibrates the baryon, lepton and DM
asymmetries.  Implementing this condition implies the following requirements: 
\begin{itemize}
\item[1.] DM must be (or must be a composite state of) a fermion, 
chiral (and thereby non-singlet) under the weak $\SU(2)_{L}$, 
and carrying an anomalous (quasi)-conserved quantum number $B'$.

\item[2.] DM (or its constituents) must have an annihilation cross section much larger
  than electroweak $\sigma_{\rm ann} \gg3~10^{-26}\cm^3/\sec$, to ensure that
  $\Omega_{\rm DM}$ is determined dominantly by the  $B'$ asymmetry.
\end{itemize}

The first condition ensures that a global quantum number corresponding to a
linear combination of $B$, $L$ and $B'$ has a weak anomaly, and thus DM
carrying $B'$ charge is produced in anomalous processes together with
left-handed quarks and leptons \cite{Kaplan:1991ah,Barr:1990ca}.  At
temperatures $T\gg T_*$ electroweak anomalous processes are in thermal
equilibrium, and equilibrate the various asymmetries $Y_{\Delta B}=c_L
Y_{\Delta L}=c_{B'}Y_{\Delta B'}\sim {\cal O}(10^{-10})$. Here the
$Y_\Delta$'s represent the difference in particle number densities $n-\bar n$
normalized to the entropy density $s$, e.g.  $Y_{\Delta B} = (n_B-\bar
n_B)/s$. These are convenient quantities since they are conserved during the
Universe thermal evolution.  

At $T\gg M_{\rm DM}$ all  particle masses can be neglected,
and $c_L$ and $c_{B'}$ are order one coefficients,  determined via chemical equilibrium conditions enforced 
by elementary reactions faster
than the Universe expansion rate~\cite{Harvey:1990qw}. These coefficients can be computed
in terms of the particle
content,  finding e.g. $c_L=-28/51$ in the SM and $c_L=-8/15$ in the MSSM.  

At $T\ll M_{\rm DM}$, the $B'$ asymmetry gets suppressed by a Boltzmann
exponential factor $e^{-M_{\rm DM}/T}$.  A key feature of sphaleron
transitions is that their rate gets suddenly suppressed at some temperature
$T_*$ slightly below the critical temperature at which $\SU(2)_L$ starts to be
spontaneously broken.  Thereby, if $M_{\rm DM}<T_*$ the $B'$ asymmetry gets
frozen at a value of ${\cal O}(Y_{\Delta B})$, while  if instead $M_{\rm DM} >
T_*$ it gets exponentially suppressed as $Y_{\Delta B'}/Y_{\Delta
  B}\sim e^{-M_{\rm DM}/T}$.

\medskip

More in detail, the sphaleron processes relate the asymmetries of the various
fermionic species with chiral electroweak interactions as follows. If
$B^{\prime}$, $B$ and $L$ are the only quantum numbers involved then the
relation is:
\begin{equation} 
\frac{Y_{\Delta B^{\prime}}}{Y_{\Delta B} } =c\cdot  {\cal S}\left(\frac{M_{\rm DM}}{T_*}\right),\qquad
c=\bar{c}_{B^{\prime}}+ \bar{c}_L \frac{Y_{\Delta L}}{Y_{\Delta B} } \ ,
\end{equation}
where the order-one $\bar{c}_{L,B^{\prime}}$ coefficients are related to the
$c_{L,B^{\prime}}$ above in a simple way.  The explicit numerical values of
these coefficients depend also on the order of the finite temperature
electroweak phase transition via the imposition or not of the weak isospin
charge neutrality. In \cite{Ryttov:2008xe,Gudnason:2006yj} the dependence on
the order of the electroweak phase transition was studied in two explicit
models, and it was found that in all cases the coefficients remain of order
one. The statistical function ${\cal S}$ is:
\begin{eqnarray}
{\cal S} (z) = \left\{ \begin{array}{rl}
\frac{6}{4\pi^2} \int_{0}^{\infty} dx\ x^2\cosh^{-2} \left( \frac{1}{2} \sqrt{x^2 + z^2 } \right) &\qquad  {\rm for~fermions} \ , \\
\frac{6}{4\pi^2} \int_{0}^{\infty} dx\ x^2\sinh^{-2} \left( \frac{1}{2} \sqrt{x^2 + z^2 } \right) &\qquad {\rm for~bosons} \ .
\end{array} \right.
\end{eqnarray}
with $S(0) = 1 (2)$ for bosons (fermions) and $S(z) \simeq 12~(z/2\pi)^{3/2} e^{-z}$ at $z\gg 1$.
We assumed the Standard Model fields to be relativistic and checked that this is a good approximation even for the top quark \cite{Gudnason:2006yj,Ryttov:2008xe}. 
The statistic function leads to the two limiting results:
\begin{equation}
\frac{Y_{\Delta B'}}{Y_{\Delta B}} =c\times  \left\{ 
\matrix{{\cal S}(0)\mbox{\hspace{2.3cm}}  &  \qquad  \hbox{for }M_{\rm DM}\ll  T_* \cr 
12\left({M_{\rm DM}}/{2\pi T_*}\right)^{3/2}\,e^{-M_{\rm DM}/T_*} & \qquad \hbox{for }M_{\rm DM}\gg T_* 
}
 \right. \quad.
\end{equation}
Under the assumption that all antiparticles carrying $B$ and $B'$ charges are
annihilated away we have $Y_{\Delta B'}/Y_{\Delta B}=n_{B'}/n_B$. The observed
DM density
\begin{equation}
\frac{\Omega_{\rm DM}}{\Omega_B}=\frac{M_{\rm DM}\,n_{B'}}{m_{p}\,n_B}\approx 5
\end{equation}
(where $m_p\approx 1\,$GeV) can be reproduced for two possible 
values of the DM mass:
\begin{itemize}
\item[i)]  $M_{\rm DM} \sim 5\,$GeV if $M_{\rm DM}\ll T_*$, times model dependent order one coefficients.
\item[ii)] $M_{\rm DM}\approx  8\, T_*\approx 2\TeV$ if $M_{\rm DM}\gg T_*$,
with a mild dependence on the model-dependent order unity coefficients.
\end{itemize}
The first solution is well known~\cite{Kaplan:1991ah} and not interesting for our purposes.
The second solution 
matches the DM mass suggested by ATIC,
in view of $T_* \sim v$~\cite{Barr:1990ca}, where $v\simeq 250\,$GeV is
the value of the electroweak breaking order parameter\footnote{More precisely,
for a Higgs  mass $m_h=120\, (300)\,$ GeV,
  Ref.~\cite{Burnier:2005hp} estimates $T_*\approx  130\, (200)\,$GeV, where the    
larger $T_*$ values  arise because of the larger values Higgs self coupling. 
For the large masses that are typical of
composite Higgs models, the self coupling is in principle 
calculable and generally large~\cite{Gudnason:2006yj}, so that 
taking $T_* \sim v$ is not unreasonable.}.

\subsection{DM mass from strongly interacting dynamics}

The ideal DM candidate suggested by the PAMELA and ATIC anomalies, and
compatible with direct DM searches,  is a $\sim$ 2 TeV particle that decays 
dominantly into leptons, and that has a negligible coupling to the $Z$.


If DM is an elementary particle, this scenario needs DM to be a chiral fermion
with $\SU(2)_L$ interactions, which is very problematic.  Bounds from direct
detection are violated.  Furthermore, a Yukawa coupling $\lambda$ of DM to the
Higgs gives the desired DM mass $M_{\rm DM}\sim \lambda v\sim 2\TeV$ if
$\lambda\sim 4\pi$ is non-perturbative, hinting to a dynamically
generated mass associated to some new strongly interacting dynamics
\cite{Nussinov:1985xr,Barr:1990ca, Ryttov:2008xe,Gudnason:2006yj}.  This
assumption also solves the problem with direct detection bounds, which are
satisfied if DM is a composite $\SU(2)_L$-singlet state, made of elementary
fermions charged under $\SU(2)_L$.

This can be realized by introducing a strongly-interacting ad-hoc `hidden'
gauge group.  A more interesting identification, namely {\it technicolor}, is
suggested by the the proximity of the DM mass indicated by ATIC, $M_{\rm
  DM}\sim 2\,$TeV, with the $4\pi v$ scale at which strong dynamics might
naturally generate the breaking of the electroweak symmetry.  In such a
scenario, DM would be the lightest (quasi)-stable composite state carrying a
$B'$ charge of a theory of dynamical electroweak breaking featuring a spectrum
of technibaryons $(B')$ and technipions ($\Pi$).

\smallskip

Let us elaborate more quantitatively on this numerical connection.
The DM mass can be approximated as $m_{B^{\prime}} = M_{\rm DM}\approx
n_Q \Lambda_{\rm TC}$ where $n_Q$ is the number of techniquarks $Q$ bounded into
$B'$ and $\Lambda_{\rm TC}$ is the constituent mass, so that $M_{\rm DM}/m_p \approx
n_Q\Lambda_{\rm TC}/3\Lambda_{\rm QCD}$.  Denoting by $f_\pi$ ($F_\Pi$) the
(techni)pion decay constant, we have $ F_\Pi/f_\pi = \sqrt{D/3}\>
\Lambda_{\rm TC}/\Lambda_{\rm QCD}$ where $D_Q$ is the dimension of the constituent
fermions representation ($D=3$ in QCD)\footnote{The large-$N$ counting
  relevant for a generic extension of technicolor type can be found in
  Appendix F of~\cite{Sannino:2008ha} together with a general introduction to
  recent models of dynamical electroweak symmetry breaking.}. Finally, the
electroweak breaking order parameter is obtained as $v^2=N_D F^2_{\Pi}$, from the sum of the
contribution of the $N_D$ electroweak techni-doublets.
Putting all together yields the estimate:
\beq
M_{\rm DM} \approx  \frac{n_Q}{\sqrt{3 D_Q N_D}} \frac{v}{f_\pi}m_p
= 2.2\TeV
\eeq
where the numerical value corresponds to the smallest number of constituents
and of techniquarks $n_Q=D_Q=2$ and $N_D=1$.  We conclude that naive rescaling
of QCD yields a value of $M_{\rm DM}$ right in the ballpark suggested by ATIC.\footnote{
Besides to the possibility indicated above, a dynamical origin of the breaking
of the electroweak symmetry can lead to several other interesting DM
candidates (see \cite{Sannino:2008ha} for a list of relevant references).}

\subsection{Phenomenological constraints on techni-DM}
It is worth mentioning that models of
dynamical breaking of the electroweak symmetry do support the possibility of 
generating the experimentally observed baryon (and possibly also the technibaryon/DM)
asymmetry of the Universe directly at the electroweak phase
transition~\cite{Cline:2008hr}.  Electroweak baryogenesis
\cite{Shaposhnikov:1986jp} is, however, impossible in the Standard Model
\cite{Kajantie:1995kf} (see \cite{Cline:2006ts} for a review on this topic).

Bounds from direct detection experiments such as CDMS and {\sc Xenon} suggest
that a single component DM candidate can have, at most, a tiny coupling with
the photon~\cite{Dimopoulos:1989hk}, the
$Z$~\cite{Caldwell:1988su,Rybka:2005vv} and the gluons.  We therefore assume
that our technibaryon $B^{\prime}$ is a singlet of $\SU(2)_L$ with zero
hypercharge (and thus also electrically neutral).  Being composed of fermions
with electroweak interactions, analogously to the neutron, the DM particles
will have electro(weak) magnetic dipoles and other similar form factors, that
can be parameterized at low energy by non-renormalizable effective operators
suppressed by powers of the technicolor scale $\Lambda_{\rm TC}$. For example,
if the $B^{\prime}$ particle is a composite scalar, the relevant operators are
suppressed at least by two powers of $\Lambda_{\rm
  TC}$~\cite{Bagnasco:1993st}, and it has been checked in~\cite{Ryttov:2008xe}
that in this case the expected interaction rate remains below the present
bounds~\cite{Ahmed:2008eu}.  An interesting collection of other
{techno-cosmology} estimates can be found in~\cite{Nussinov:1985xr}.

\subsection{DM lifetime and decay modes}
According to~\cite{Rubakov} the sphaleron contribution
to the  techni-baryon decay rate is negligible because exponentially
suppressed, unless the techni-baryon is heavier than several TeV.

Grand unified theories (GUTs) suggest that the baryon number $B$ is violated
by dimension-6 operators suppressed by the GUT scale $M_{\rm GUT} \sim2\cdot
10^{16}\GeV$, yielding a proton  life-time~\cite{Hisano:1992jj}
\beq
\tau(p\to \pi^0 e^+) \sim \frac{M_{\rm GUT}^4}{m_p^5}\sim
10^{41} \,{\rm sec}.
\eeq
If $B'$ is similarly violated at the same high scale $M_{\rm GUT}$,
our DM techni-baryon would decay with life-time
\beq 
\tau \sim \frac{M_{\rm GUT}^4}{M_{\rm DM}^5}\sim 10^{26}\,{\rm sec}, 
\eeq
which falls in the ball-park required by the phenomenological analysis above.
Models of unification of the Standard Model couplings in the
presence of a dynamical electroweak symmetry breaking mechanism have been
recently explored \cite{Christensen:2005bt,Gudnason:2006mk}. Interestingly, the
scale of unification suggested by the phenomenological analysis emerges quite
naturally \cite{Gudnason:2006mk}.

\smallskip 

Low energy DM and nucleon (quasi)-stability imply that, in the primeval
Universe, at temperatures $T\circa{<} M_{\rm GUT}$ perturbative violation of
the $B'$ and $B$ global charges is strongly suppressed. Since this temperature
is presumably larger than the reheating temperature, it is unlikely that
$\Omega_{B}$ and $\Omega_{\rm DM}$ result directly from an asymmetry generated 
in $B'$ or  $B$. More likely, the initial seed yielding $\Omega_{\rm DM}$ and $\Omega_B$
could be an initial asymmetry in lepton number $L$ that, much along the lines 
of well studied leptogenesis scenarios~\cite{fu86},  feeds the $B$ and $B'$ asymmetries 
through the sphaleron effects.\footnote{In Minimal Walking Technicolor \cite{Dietrich:2005jn,Foadi:2007ue}, one additional  (techni-singlet)  $\SU(2)$-doublet  must be introduced 
to cancel the odd-number-of-doublets anomaly~\cite{Witten:1982fp}. 
An asymmetry in the $L'$ global charge associated with these new states can 
also serve as a seed for the $B$ and $B'$ asymmetries.}
Indeed, it has been shown that it is possible to embed seesaw-types of
scenarios in theories of dynamical symmetry breaking,
while keeping the scale of the $L$-violating Majorana masses as low as 
$\sim 10^{3}\,$TeV~\cite{Appelquist:2002me}. 

\medskip

Assuming that techni-baryon DM decay is dominantly due to effective four-fermion operators,
its decay modes significantly depend on the technicolor gauge group.
In the following $F$ generically denotes any SM fermion, quark or lepton,
possibly allowed by the Lorentz and gauge symmetries of the theory.
\begin{itemize}
\item If the technicolor group is SU(3), the situation is  analogous to ordinary QCD:
DM is a fermionic $QQQ$ state, and effective $QQQF$ operators
gives $\DM\to \Pi^- \ell^+ $ decays.  
This leads to hard leptons, but together with an excess of $\bar p$,
from the $\Pi^- \to s \bar c$ decay (in view of  $\Pi^- \simeq  W^-_L$).  
This is therefore incompatible with PAMELA $\bar p$ data, unless astrophysical propagation uncertainties are relaxed.
\item 
If the technicolor group is SU(4) the situation is typically worse:
DM is a bosonic $QQQQ$ state, and effective $QQQQ$ operators lead to
its decay into techni-pions.
\item Finally, if the technicolor gauge group is SU(2), DM is a bosonic $QQ$
  state, (as put forward in \cite{Ryttov:2008xe}), and effective $QQFF$
  operators lead to DM decays into two $F$.  Since the fundamental
  representation of SU(2) is pseudoreal, one actually gets an interesting
  dynamics analyzed in detail in \cite{Ryttov:2008xe}.  Here the technibaryon
  is a pseudo-Goldstone boson of the underlying gauge theory.
\end{itemize}
An SU(2) technicolor model compatible with the desired features is obtained
assuming that the left component of the Dirac field $Q$ has zero hypercharge
and is a doublet under $\SU(2)_L$, so that DM is a scalar $QQ$ with no weak
interactions, and the four-fermion operator $(QQ)\partial_\mu (\bar F
\gamma_\mu F)$ allows it to decay.  Such operator is possible for both SM
leptons and quarks, so that the DM branching ratios into $\ell^+ \ell^-$ and
$q\bar q$ is a free parameter.

%


\smallskip 

The use of higher dimensional representations for the techniquarks
transforming under a given technicolor gauge group has opened new
possibilities \cite{Sannino:2008ha}. Interestingly, the technicolor model with
the lowest possible value for the naive $S$ parameter (measuring deviations
from electroweak precision data) can be constructed by combining {\it exactly}
two Dirac flavors in the fundamental of $\SU(2)$ with one Dirac in the adjoint
representation (uncharged with respect to the SM) \cite{Ryttov:2008xe}.  The
model has also the lowest number of fermions and is compatible with near
conformal dynamics, which is an important ingredient for reducing the tension
with constraints stemming from unobserved flavor changing neutral currents.

\section{Conclusions}
We explored the interpretation of the excess of $e^\pm$ cosmic rays observed by
PAMELA and ATIC in terms of decays of Dark Matter with mass $M_{\rm DM}$.

On the phenomenological side, we found that any DM decay modes involving hard
leptons can fit the PAMELA excess for $M_{\rm DM}\circa{>}200\GeV$, while decay
modes that give soft leptons with energy $E\ll M_{\rm DM}$ together with $\bar p$
 need $M_{\rm DM}\circa{>} 10\TeV$.  The peak  present around 800 GeV in the
$e^++e^-$ ATIC spectrum can be reproduced if $M_{\rm DM}\sim 2\TeV$ and for DM
decays into $\mu^+\mu^-, \tau^+\tau^-$ (characteristic of boson DM) and into
$W^\pm \ell^\mp$ where $\ell$ is any lepton (characteristic of fermion DM).
However, the latter decays are disfavored by the PAMELA non-observation of an
excess in the $\bar p$ CR spectrum.  Fermionic DM is allowed if it
decays into $\ell^+\ell^-\nu$.

DM annihilations compatible with the PAMELA excess tend to give a flux of
$\gamma$ rays above the HESS~\cite{HESSSgrDwarf,HESSGC, HESSGR} bounds and a
flux of synchrotron radiation above the Davies~\cite{Davies} bounds, unless
the DM density profile $\rho(r)$ is significantly less steep as $r\to 0$
than the NFW profile.
Whether or not this should be considered a problem depends on the
the reliability of $N$-body simulations, and of their extrapolation to scales below 100 pc.

\medskip

Anyhow, this issue is not present for DM decays, because they give DM signals proportional to $\rho$
rather than to $\rho^2$.  The HESS observation of the Galactic Ridge region
gives the dominant constraint, which is just below the needed sensitivity, and
can be improved by performing an observation optimized for DM
rather than for astrophysics.

On the theoretical side, we found that it is easy to invent decaying DM models
compatible with the above features suggested by data, just because
model-building is less severely constrained than in the case of DM
annihilations. 

More interestingly, if the DM density is due to an asymmetry analogous to
baryon number broken at low energy by weak anomalies, then the value of the DM
mass suggested by the $e^++e^-$ peak, $M_{\rm DM} \sim2\TeV$, is also the one that
naturally gives the observed ratio between the dark and $b$aryonic matter
densities $\Omega_{\rm DM}/\Omega_B \sim 5$.

Indeed, in such a case electroweak sphalerons keep $n_{\rm DM}\sim n_B$ down
to $T \circa{>} M_{\rm DM}$; below this temperature the DM density gets
Boltzmann-suppressed, until sphalerons freeze-out at the temperature
$T_{*}\sim 200\GeV$.  As a result, the correct DM density is obtained for $M
\sim 8 T_{*}$, which coincides with the mass suggested by the ATIC peak.
Technicolor extensions of the SM suggest a similar value for the mass of the
lightest (quasi)-stable state.  Technicolor is a new strongly coupled gauge
dynamics which breaks the electroweak symmetry dynamically and solves the
hierarchy problem without an elementary Higgs.  These models feature
technibaryons which are quasi-stable technihadrons analogous to the proton in
QCD, that possess, quite surprisingly, many of the required properties
advocated by our phenomenological analysis.

While these phenomenological coincidences are indeed suggestive, from the
theoretical point of view, bound states of some new strong dynamics decaying predominantly into leptons require some specific construction, that seems to privilege an SU(2) technicolor gauge theory with a technimatter structure envisioned in \cite{Ryttov:2008xe}.

\medskip

\bigskip

\footnotesize

\begin{multicols}{2}

\end{multicols}

\end{document}